\documentclass[amsmath,amssymb,reprint,superscriptaddress]{revtex4-1}

\usepackage{graphicx} 
\usepackage{hyperref}
\usepackage{lipsum}
\usepackage{color} 
\usepackage{lmodern} 
\usepackage{textcomp} 
\usepackage[T1]{fontenc} 

\DeclareMathOperator*{\Trace}{Tr}
\newcommand*{\ket}[1]{|{#1}\rangle}
\newcommand*{\bra}[1]{\langle{#1}|}

\newcommand*{\mean}[1]{\mathinner{\langle{#1}\rangle}}

\def\be{\begin{equation}}
\def\ee{\end{equation}}
\def\bes{\begin{equation*}}
\def\ees{\end{equation*}}

\newcommand{\parttl}[1]{}

\newcommand{\bout}{\hat b_{\rm out}}

\begin{document}

\title{Two-Photon Resonance Fluorescence of a Ladder-Type Atomic System}

\author{Simone~Gasparinetti}
\email{simoneg@chalmers.se}
\homepage[Current address: ]{Department of Microtechnology and Nanoscience MC2, Chalmers University of Technology, Kemiv\"agen 9, SE-41296 G\"oteborg, Sweden}
\affiliation{Department of Physics, ETH Zurich, CH-8093 Zurich, Switzerland}

\author{Jean-Claude~Besse}
\affiliation{Department of Physics, ETH Zurich, CH-8093 Zurich, Switzerland}

\author{Marek~Pechal}
\affiliation{Department of Physics, ETH Zurich, CH-8093 Zurich, Switzerland}

\author{Robin~D.~Buijs}
\affiliation{Department of Physics, ETH Zurich, CH-8093 Zurich, Switzerland}

\author{Christopher~Eichler}
\affiliation{Department of Physics, ETH Zurich, CH-8093 Zurich, Switzerland}

\author{Howard~J.~Carmichael}
\affiliation{The Dodd-Walls Centre for Photonic and Quantum Technologies, Department of Physics,
University of Auckland, Private Bag 92019, Auckland, New Zealand}

\author{Andreas~Wallraff}
\affiliation{Department of Physics, ETH Zurich, CH-8093 Zurich, Switzerland}

\date{\today}

\begin{abstract}
Multi-photon emitters are a sought-after resource in quantum photonics. Nonlinear interactions between a multi-level atomic system and a coherent drive can lead to resonant two-photon emission, but harvesting light from this process has remained a challenge due to the small oscillator strengths involved.
Here we present a study of two-photon resonance fluorescence at microwave frequencies, using a superconducting, ladder-type artificial atom, a transmon, strongly coupled to a waveguide. We drive the two-photon transition between the ground and second-excited state at increasingly high powers and observe a resonance fluorescence peak whose intensity becomes comparable to single-photon emission until it  splits into a Mollow-like triplet. We measure photon correlations of frequency-filtered spectral lines and find that while emission at the fundamental frequency stays antibunched, the resonance fluorescence peak at the two-photon transition is superbunched. Our results provide a route towards the realization of multi-photon sources in the microwave domain.
\end{abstract}

\maketitle

\begin{figure}
\includegraphics[width=\linewidth]{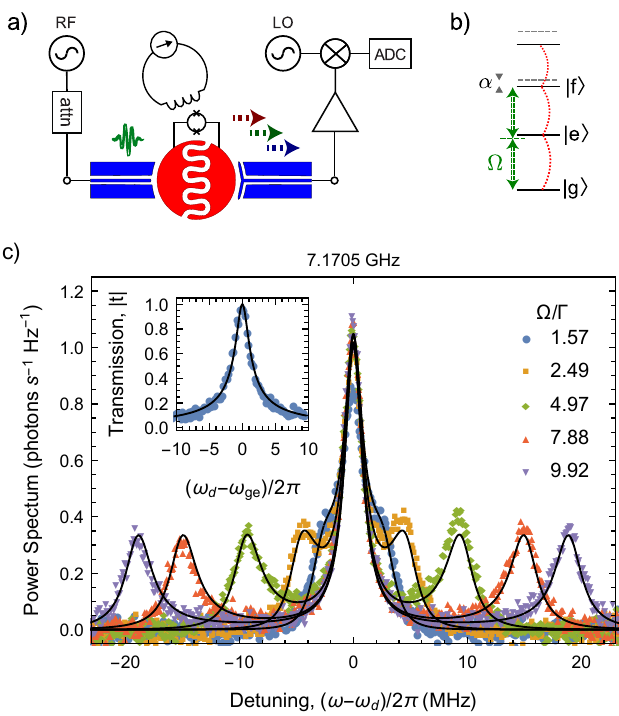}
\caption{\textbf{Ladder-type artificial atom strongly coupled to a waveguide.}
(a) Simplified scheme of the experimental setup. The transmon consists of a planar capacitor (red) shunted by a nonlinear inductor, a SQUID, which can be tuned by static magnetic field, and it is weakly (strongly) coupled to an input (output) port. The input is driven by an attenuated, coherent microwave source (RF). 
The output is amplified using a chain of linear amplifiers, mixed with a local oscillator (LO) 
and recorded with a fast digitizer (ADC).
(b) Level scheme for the first three levels of the transmon, $\ket{g}$, $\ket{e}$, and $\ket{f}$, with anharmonicity $\alpha$. Red, dotted lines indicate the allowed first-order transitions. The second-order, two-photon transition between $\ket{g}$ and $\ket{f}$ is resonantly driven with strength $\Omega$ (green, double lines).
(c) Measured inelastic scattering spectrum of the transmon driven at its fundamental frequency, $\omega_{ge}$, and at the indicated Rabi rates, $\Omega$, normalized to the decay rate, $\Gamma$ (symbols). The black lines are a global fit to the data.
Inset: low power spectroscopy of the transmon (dots) and Lorentzian fit (red line) used to extract the resonant frequency, $\omega_{ge}$ and the linewidth, $\Gamma$.
}
\label{fig:setup}
\end{figure}

\parttl{Resonance fluorescence} Resonance fluorescence, the resonant scattering of electromagnetic radiation from a two-level atom or molecule, provides a signature of coherent light-matter interactions as well as a source of nonclassical radiation, and is therefore regarded as a cornerstone of quantum optics.
Upon strong, coherent, resonant irradiation, the fluorescence spectrum of the atom exhibits a Mollow triplet \cite{Mollow1969,Schuda1974,Wu1975}, which is interpreted as radiative decay down a ladder of dressed states \cite{Cohen-Tannoudji1998}. The emitted radiation is strongly anti-bunched \cite{Kimble1977} and can be utilized as a source of single photons \cite{Eisaman2011}. Various features of resonance fluorescence have been recently explored using semiconductor quantum dots, including cascaded photon emission from the triplet sidebands \cite{Ulhaq2012}, sideband enhancement by coupling to a cavity \cite{Kim2014e}, suppression of the resonance fluorescence line by a bichromatic drive \cite{He2015}, and correlations between photons of different color \cite{Peiris2015}. Resonance fluorescence has also been studied in nonlinear superconducting circuits (``artificial atoms'') with transition frequencies in the microwave range, which can be strongly coupled to well-controlled, itinerant modes of one-dimensional waveguides \cite{Astafiev2010,Lang2011,Hoi2012b,vanLoo2013,Toyli2016a}.

\parttl{Motivation for TPRF} In two-photon fluorescence, the emitter is excited by two-photon absorption,
a technique that
has found application, for instance, in the imaging of biological samples \cite{Denk1990}. In optical systems, two-photon transitions typically occur between states of the same parity, and can be assisted by -- one or more, resonant or off-resonant -- intermediate states of opposite parity.
In the presence of a dipole-coupled intermediate state, the fluorescence spectrum is dominated by the emission of two photons in cascade. This scenario has been recently explored in the biexciton-exciton cascade of semiconductor quantum dots \cite{Ardelt2016,Hargart2016,Bounouar2017}, as well as in superconducting circuits \cite{Gasparinetti2017}. At the same time, it has been shown theoretically that quasi-resonant intermediate states can assist inelastic scattering at the drive frequency, giving rise to two-photon \textit{resonance} fluorescence \cite{Alexanian2006}. The experimental characterization of this effect has so far proven difficult, possibly due to the presence of competing decay channels. In Ref.~\cite{Hargart2016}, two-photon resonance fluorescence was strongly enhanced by embedding the quantum dot into a cavity resonant with the two-photon transition (Purcell effect). However, the scattered light of the strong pump laser field saturated the detector and thereby impeded a clean measurement.
If the two-photon signal can be separated from the coherent background,
it may serve as a source of non-classical light. This type of multi-photon emitters  are expected to have applications in quantum photonics \cite{Munoz2014}, quantum metrology \cite{Giovannetti2006},
and quantum biology \cite{Mohseni2014,Potocnik2018}.

\parttl{Experiment in context} Here we present an experimental
study of two-photon resonance fluorescence in the microwave frequency domain, using a superconducting circuit, a transmon \cite{Koch2007}, as a ladder-type artificial atom. We measure the spectrum of inelastic scattering for coherent pump strengths ranging from much weaker up to comparable to the intermediate-state detuning, as well as photon-photon correlations of individual, frequency-filtered spectral lines corresponding to one- and two-photon transitions.
Our results are in good agreement with a theory based on three-level dressed states \cite{Koshino2013}, analytical expressions for the decay operators corresponding to different dressed-state transitions, master equation and quantum regression theorem.
Remarkably, the setting explored here is closely related to two-photon blockade in a cavity-QED system \cite{Hamsen2017}, with the anharmonicity of the transmon replacing the nonlinear splitting of the bare cavity states as they resonantly interact with the atom. As a matter of fact, emission spectra qualitatively similar to those presented here have been predicted for doubly-dressed states in cavity QED \cite{Shamailov2010}.

\parttl{Setup} The device under study consists of a transmon emitting radiation towards a microwave switch embedded on the same chip, and was characterized in previous work \cite{Pechal2016,Gasparinetti2017}.
The transmon is asymmetrically coupled to two waveguides [Fig.~1(a)]: a weakly coupled drive line, used for coherent excitation, and a strongly coupled output line. 
This arrangement makes it possible to apply a strong drive to the transmon with minimal leakage into the output line \cite{Peng2016d,Pechal2016,Pechal2016a}, and to collect a large fraction ($\approx 98\%$) of the radiation emitted by the transmon.
The output is amplified by an amplification chain having a Josephson parametric dimer (JPD) operated in the phase-insensitive mode as the first amplifier \cite{Eichler2014a}. The JPD has tunable gain, center frequency, and bandwidth. In order to maximize the signal-to-noise ratio and minimize spurious effects such as nonlinearities and inhomogeneities in the density of states seen by the transmon at each transition frequency, the data presented in this manuscript was taken at the transmon/JPD configurations summarized in the Supplementary Materials \cite{SM}. 

\parttl{Transmon characterization} The frequency spectrum of the transmon [Fig.~1(b)] is that of a nonlinear oscillator \cite{Koch2007} with negative anharmonicity $\alpha/2\pi=-233~\rm{MHz}$. We measure the fundamental frequency $\omega_{ge}/2\pi$ by low-power transmission spectroscopy [Fig. 1(c), Inset] and tune it by static magnetic field in a range between $7.180~\rm{GHz}$ and $7.500~\rm{GHz}$. The corresponding linewidth, $\Gamma/2\pi$, varies between $1.9~\rm{MHz}$ and $3.4~\rm{MHz}$ in the range considered.
We ascribe this variation to phase interference effects due to impedance mismatches in the microwave hybrids composing the switch \cite{Pechal2016}, which in this work is only used to forward the radiation emitted by the transmon to the output line.
Resonant driving of the $g$--$e$ transition produces the well-known Mollow triplet [Fig.~1(c)] \cite{Astafiev2010, Lang2011}. A global fit of the emission spectra \emph{vs} power is used to calibrate the Rabi rate, $\Omega$, as a function of input power, $P_{\rm in}$, assuming the expected linear dependence of the form $\Omega \propto \sqrt{P_{\rm in}}$.

\begin{figure*}%
\includegraphics[width=\linewidth]{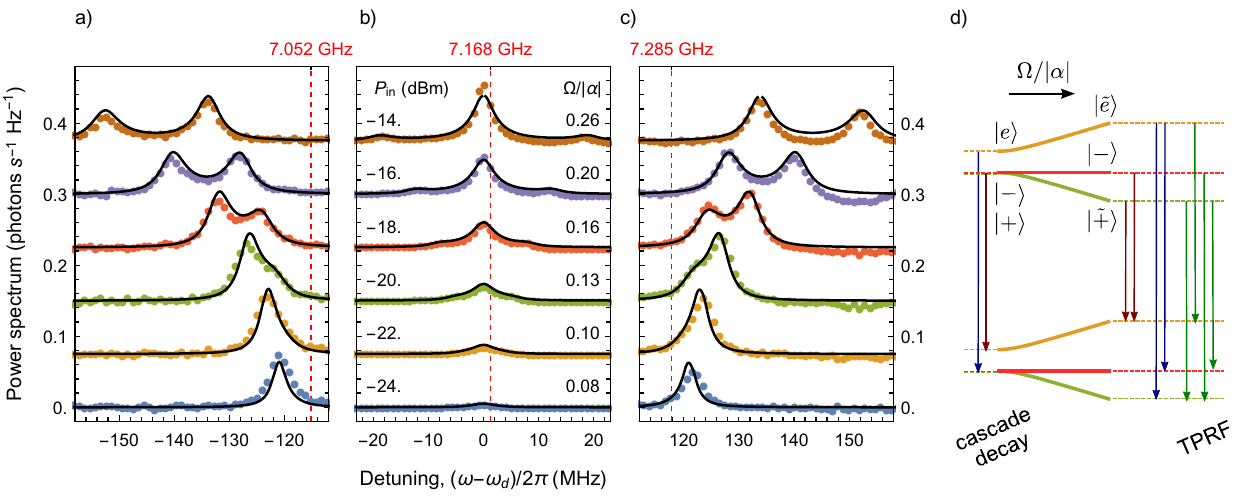}
\caption{\textbf{Emission power spectra under strong two-photon excitation.} (a-c) Measured power spectral density versus detuning from the drive frequency, $\omega-\omega_d$, in the vicinity of (a) the $e$--$f$, (b) the two-photon $g$--$f$, and (c) the $g$--$e$ transition, for the indicated input powers $P_{\rm in}$ (dots). Bare transition frequencies are indicated by red dashed lines and the traces are vertically offset for clarity. The black lines are obtained by a numerical simulation (see main text for details).
(d) Level scheme illustrating inelastic transitions in the dressed-state basis $\ket{\tilde{e}},\ket{-},\ket{\tilde{+}}$, in the limits of low ($\Omega/|\alpha| \ll 1$, left) and high drive power ($\Omega/|\alpha| \lesssim 1$, right), corresponding to cascade decay and two-photon resonance fluorescence (TPRF).
}
\label{fig:PSDs}%
\end{figure*}

\parttl{Power spectrum}
We drive the transmon at the two-photon transition $\omega_{gf}/2$ and measure the emission spectrum at frequencies close to the relevant transitions at $\omega_{ge}$, $\omega_{ef}$, and $\omega_{gf}/2$, for different input powers [Fig.~2(a-c), dots]. The JPD is configured so that it amplifies the signal around $\omega_{gf}/2$,
while the signal from other transitions is simply directed to the next amplifier in the chain.
At low power, most of the emission results from the two single-photon transitions, as expected for cascade decay \cite{Gasparinetti2017}, while a comparatively smaller inelastic scattering peak appears at the drive frequency, corresponding to two-photon resonance fluorescence. At higher powers, each of the two single-photon peaks is Stark-shifted away from the two-photon transition, and split into a doublet \cite{Ardelt2016,Hargart2016,Bounouar2017}. At the same time, the two-photon peak grows as strong as the single-photon peaks, and eventually splits into a triplet. The separation between the triplet sidebands and the central peak is similar to the separation between each doublet, and scales approximately linearly with the applied power.
The data is presented along with a numerical simulation including the first four levels of the transmon (black lines). The simulation assumes a constant density of states for the waveguide and a $\sqrt{n}$-scaling for the oscillator strength of the transition between the levels $n$ and $(n+1)$ \cite{Koch2007}. The parameters of the simulation are the fundamental frequency of the transmon, $\omega_{ge}$, its  anharnmonicity, $\alpha$, the $g$--$e$ linewidth, $\Gamma$, and the total attenuation of the input line. These parameters are all extracted from independent spectroscopy measurements, with the exception of the linewidth, for which the value $\Gamma/2\pi=2.5~\rm{MHz}$ was chosen as an interpolation between the measured $g$--$e$ transition linewidths across the frequency range of the three transitions \cite{SM}. We notice that including the fourth level of the transmon does not change the qualitative description of the effect but it is needed to obtain a quantitative agreement with the measured spectra at the highest powers \cite{SM}.

\parttl{Explanation: dressed-state ladder}
We explain the observation of seven emission peaks by considering a ladder of dressed states for the first three states of the transmon [Fig.~2(d)].
When the drive strength is much lower than the anharmonicity, $\Omega \ll |\alpha|$, the states $\ket{g}$ and $\ket{f}$, degenerate in a frame rotating at the drive frequency, hybridize to form the quasidegenerate pair $\ket{-}$ and $\ket{+}$, while the intermediate state $\ket{e}$ is not dressed. In this setting, cascade decay (from $\ket{+}$ to $\ket{e}$ and then from $\ket{e}$ to $\ket{+}$) is the only allowed relaxation path \cite{Gasparinetti2017}. Instead, for drive strengths comparable to the anharmonicity, $\ket{e}$ hybridizes with $\ket{+}$ to form the three-state dressed states $\ket{\tilde{+}}$ and $\ket{\tilde{e}}$ \cite{Koshino2013}. This hybridization causes a repulsion of the respective dressed-state energies (by an amount $3\Omega^2/|\alpha|$, at resonance and up to second order in $\Omega/|\alpha|$), which can be observed as a shift and a splitting of the two single-photon peaks [Fig.~2(a,c)]. Even more importantly, three-state dressing allows for additional relaxation paths which entail emission around the drive frequency, and are therefore responsible for the strong signal observed at the two-photon transition in Fig.~2(b).

\parttl{Correlation function measurements}
We investigate the nonclassical character of the emitted radiation by measuring the normalized second-order correlation function, $g_2(\tau)$. We perform power autocorrelation measurements of the signal emitted at the single-photon transition, $\omega_{ge}$, and at the two-photon transition, $\omega_{gf}/2$, using the techniques described in Refs.~\cite{Bozyigit2011,daSilva2010,Eichler2012}. In our setup, the limited bandwidth of the JPD naturally acts as a frequency filter. For each spectral line to be analyzed, we tune the frequency of the transmon so that the emission is peaked at the center of our detection window.
We further restrict the detection bandwidth to $\sim 12~\rm{MHz}$ by digital filtering, and vary the input power in a range in which (i) most of the emitted radiation around each transition lies within the detection window, and (ii) the fourth level of the transmon is not expected to be involved in the dynamics (based on theory simulations).
We find radiation around $\omega_{ge}$ to be antibunched ($g_2(0) \ll 1$), with a characteristic time scale that becomes shorter with increasing drive power [Fig.~3(a)]. By contrast, radiation around $\omega_{gf}/2$ displays superbunching ($g_2(0)>1$). The magnitude of photon correlations increases with decreasing drive strength [Fig.~3(b)].

\begin{figure}[t]
\includegraphics[width=\linewidth]{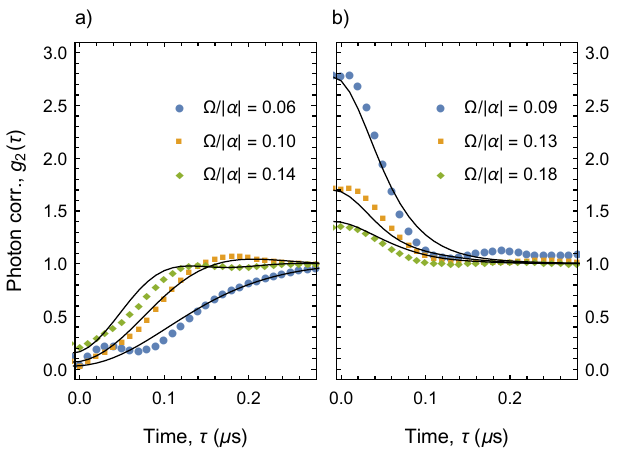}
\caption{\textbf{Photon correlations of radiation emitted at different transitions.}
Normalized power autocorrelation function, $g_2(\tau)$ for frequency-filtered radiation emitted at the frequencies (a) $\omega_{ge}$ and (b) $\omega_{gf/2}$ upon resonant driving $\omega_{gf}/2$ at the indicated drive powers (symbols). For both datasets, the amplification chain is centered at $7.358~\rm{GHz}$ and has a bandwidth of 12 MHz. The transmon fundamental frequency, $\omega_{ge}/2\pi$ is adjusted by static magnetic flux to (a) $7.352~\rm{GHz}$ and (b) $7.477~\rm{GHz}$, so that emission at the relevant transition falls within the acquisition band.
Solid lines are the corresponding calculated $g_2(\tau)$ (see text and \cite{SM} for details).
}
\label{fig:g2}
\end{figure}

\parttl{Modeling of frequency-filtered correlations}

We calculate photon-photon correlations using input-output-theory, master equation, and the quantum regression theorem \cite{Carmichael2002}. To compare our model to the measurements of Fig.~3(a,b), we need to calculate frequency-filtered photon correlations \cite{delValle2012} for each individual spectral line.
To do so, we explicitly decompose the total field operator into a sum of three operators accounting for transitions at the frequencies $\omega_{ge}$, $\omega_{ef}$, and $\omega_{gf}/2$, respectively. These field operators can be treated independently after invoking a secular approximation in the dressed-state basis \cite{SM}. For this analysis we have restricted ourselves to the first three levels of the transmon.

\parttl{Comparing to experiment}

We find the predictions of our theory [solid lines in Fig.~3(a,b)] to be in qualitative agreement with the experimental data.
In particular, our model captures the crossover between superbunching and Poissonian statistics ($g_2(0)=1$) observed for the two-photon resonance fluorescence signal when increasing the drive strength [Fig.~3(b)]. The availability of an explicit expression for the corresponding field operator can be used to gain insight on the mechanism underlying the crossover. By direct inspection of the matrix elements
we identify four relaxation paths connecting dressed-states with different photon numbers which contribute to resonance fluorescence, namely, $\ket{-} \to \ket{\tilde{+}}$, $\ket{\tilde{+}} \to \ket{-}$, $\ket{\tilde{+}} \to \ket{\tilde{+}}$, and $\ket{\tilde{e}} \to \ket{\tilde{e}}$. Of these, the first two are expected to have two-photon character because they connect the bare states $\ket{g}$ and $\ket{f}$, whose energies differ by two photons at the two-photon resonance. By contrast, the last two relaxation paths do not entail population modulation of the state of the emitter. For these paths, subsequent emission events are uncorrelated and the photon statistics can thus be expected to be Poissionian;
a similar phenomenology is observed for the central peak of the standard Mollow triplet \cite{Schrama1992,Ulhaq2012}.
At small drive strengths, the steady-state solution of the master equation indicates that the dressed-state $\ket{-}$ is the most populated, which leads to superbunching in two-photon fluorescence. For increasing drive strength, emission at the two-photon transition intensifies due to the increased participation of the intermediate state $\ket{e}$. At the same time, however, the occupation of the dressed-state $\ket{\tilde{e}}$ grows at the expense of $\ket{-}$, favoring Poissonian decay and thereby decreasing the purity of two-photon emission, as seen experimentally by the decrease in $g_2(0)$. Further numerical studies \cite{SM} indicate 
that by adjusting the drive detuning it should be possible to engineer the steady-state populations so that strong superbunching is achieved also at intermediate drive strengths.

\parttl{Conclusions}
In summary, we have experimentally
characterized the radiation emitted by two-photon resonance fluorescence assisted by a quasi-resonant intermediate state, using a superconducting circuit with transition frequencies in the microwave domain. We observe strong superbunching in this setting, in agreement with our model.
The pumping scheme used here could be applied to cavity-QED systems exhibiting two-photon blockade \cite{Hamsen2017}, as well as to optically active semiconductor quantum dots driven at the two-photon biexciton transition \cite{Ardelt2016}. For these systems, we expect qualitatively similar energy spectra \cite{Shamailov2010} and correlation properties, with differences possibly arising from the presence of a pair of intermediate states
(non-degenerate in the former case, quasi-degenerate in the latter).
For optical systems, the polarization properties of the emitted radiation could be subject of further study. The intensity of two-photon emission can be enhanced by embedding the emitter in a cavity resonant with the two-photon transition, as suggested in \cite{Valle2011} and realized in \cite{Hargart2016}. In this arrangement, emission at the two-photon frequency becomes the dominant relaxation channel, and one can take advantage of the high-fidelity state preparation available, e.g., for superconducting circuits to make the process deterministic. This would turn the device into an on-demand source of two-photon wavepackets, with the further possibility to use the intermediate state to shape the temporal envelope of the emitted photons \cite{Pechal2014,Kurpiers2018}.

\textit{Acknowledgements.} We are grateful to M.~Collodo and A.~Poto\v{c}nik for useful discussions.
This work was supported by the National Centre of Competence in Research ``Quantum Science and Technology'' (NCCR QSIT), a research instrument of the Swiss National Science Foundation (SNSF), and by ETH Zurich.

\nocite{Lang2014}

\appendix

\section{Experimental details}

\subsection{Working points for the transmon and Josephson parametric amplifier}

We refer to Refs.~\cite{Pechal2016} and \cite{Gasparinetti2017} for a detailed description and characterization of the sample. Due to experimental constraints, the data presented in this work were taken at various configuration of the transmon and the Josephson parametric amplifier used to detect the output radiation. The parametric amplifier was used at two selected working points (JPA\#1, JPA\#2), with measured specifications as in Table~\ref{tblWP}.

\begin{figure}[b]
\includegraphics{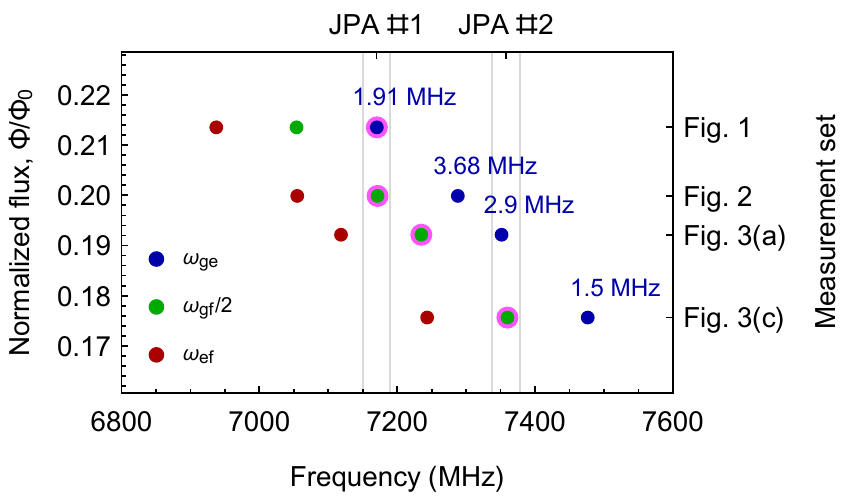}
\caption{Measured transition frequencies of the transmon ($\omega_{ge}$, blue, $\omega_{ef}$, red, $\omega_{gf}/2$, green, bottom axis) \textit{vs} normalized magnetic flux, $\Phi/\Phi_0$  (left axis) for each dataset shown in the main text (right axis). For each configuration, the driven transition (purple circle) and the measured $g$--$e$ linewidth are indicated. The lower and the upper edge of the detection band of the parametric amplifier are indicated by vertical lines for the two working points used Fig.~1 and 2 (``JPA\#1'') and in Fig.~3 (``JPA\#2'', top axis).
}
\label{Sfig:wps}
\end{figure}

For each experiment, the transmon frequency was adjusted by static magnetic flux so that the relevant transition lay in the detection bandwidth of the amplifier. A graphical summary of the settings used to produce the data is presented in Fig.~\ref{Sfig:wps}.

\bigskip

\begin{table}[b]%
\begin{tabular}{c|c|c|c}
Working point & Center frequency & Gain & 3 dB bandwidth \\
\hline
\#1 & $7.170~\rm{GHz}$ & $7~\rm{dB}$ & $56~\rm{MHz}$ \\
\hline
\#2 & $7.358~\rm{GHz}$ & $11~\rm{dB}$ & $25~\rm{MHz}$ \\
\hline
\end{tabular}
\caption{Working points for the Josephson parametric amplifier}
\label{tblWP}
\end{table}

\section{Theoretical model}\label{sec:analysis}

\subsection{Hamiltonian, master equation and input-output theory}

In a frame rotating at the drive frequency $\omega_d$ and using a rotating-wave approximation, we describe the driven transmon by the Hamiltonian
\be
\hat H= -(\delta+\alpha/2) \hat b^\dagger \hat b + \frac{\alpha}2 \hat b^\dagger \hat b^\dagger \hat b \hat b
+  i \frac{\Omega}{2} (\hat b-\hat b^\dagger) \ ,
\ee
where $\hat b$ is an annihilation operator, $\delta=\omega_d-\omega_{gf}/2$ is the drive detuning from the two-photon transition, $\alpha$ is the transmon anharmonicity ($\alpha<0$), and $\Omega$ is the drive strength \cite{Koch2007}.

We model the driven-dissipative dynamics of the system by the action of the Liouvillian $\mathcal L$ on the density matrix $\hat \rho(t)$:
\be
\mathcal L \hat\rho(t) = -\frac{i}{\hbar}[\hat H,\hat\rho(t)]+ \Gamma \mathcal D[\hat b,\hat\rho(t)]
\ ,  \label{eq:liou}
\ee
where
$\mathcal D[A,B] = A BA^\dagger - \frac12\left(A^\dagger A B- B A^\dagger A\right)$,
and $\Gamma$ is the radiative decay rate of the transmon into the transmission line \cite{Carmichael2002}.
We have neglected any other decay and relaxation channels \cite{Pechal2016}. For simplicity, we have also assumed that the transmission line has a constant density of states over the full emission spectrum.

We solve the dynamics by truncating the Hilbert space of the transmon up to $n$ excitations ($n=3,4$ are considered here). The photonic field at the output of the transmission line, described by the operator $\hat b_{\rm out}$, is related to the transmon decay operator by
\be
\hat b_{\rm out} = \sqrt{\Gamma} \hat b \ .
\ee

\subsection{Field operators describing the radiation emitted at individual transitions}

We decompose the operator $\hat b$ into a sum of operators accounting for the two single-photon doublets and for the two-photon triplet observed in the power spectrum [Fig.~2(a-c)]. We do this explicitly for the case $n=3$. First we diagonalize the Hamiltonian to find the three-level dressed states $\ket{\xi_i}$ and their eigenfrequencies $\lambda_i$.


At resonance with the bare two-photon transition, the dressed-state frequencies are
\begin{align}
\lambda_1 &= 0 \ , \\
\lambda_2 &= \frac{1}{4} \left(-\sqrt{\alpha ^2+12 \Omega ^2}-\alpha \right) \approx -\frac{3 \Omega^2}{2|\alpha|} \ , \\
\lambda_3 &= \frac{1}{4} \left(+\sqrt{\alpha ^2+12 \Omega ^2}-\alpha \right) \approx -\alpha/2 + \frac{3 \Omega^2}{2|\alpha|}    \ ,
\end{align}
where the approximate expressions hold up to second order in the ratio $\Omega/|\alpha|$. In general, $\lambda_1$ and $\lambda_2$ are quasi-degenerate (corresponding to the hybridization of the states $\ket{1}$ and $\ket{3}$ in the rotating frame), while $\lambda_3$ is detuned by approximately $|\alpha|/2$ (corresponding to the state $\ket{2}$). The states $\ket{\lambda_{1,2,3}}$ are denoted in the main text as $\ket{-}$, $\ket{\tilde{+}}$, and $\ket{\tilde{e}}$, respectively.



By expressing the bare states $\ket{i}$ in terms of the dressed states as $\ket{i} = \sum_{j} c_{ij} \ket{\xi_j}$, we write down the annihilation operator $\hat b = \sqrt{2}\ket{2}\bra{3}+\ket{1}\bra{2}$ as a sum of terms involving pairs of dressed states. In the interaction picture, each such term oscillates as the difference between two dressed-state frequencies. We group terms which oscillate at the same frequency together, and drop fast-oscillating terms by invoking a secular approximation. By selecting those terms that oscillate at frequencies much smaller than the anharmonicity, we arrive at the two-photon resonance fluorescence operator
\be
\begin{split}
\hat T  = &\left(c_{11} \ket{\xi_1} c_{12} \ket{\xi_2}\right) c_{22} \bra{\xi_2} + c_{13} c_{23}\ket{\xi_3} \bra{\xi_3} \\
& +  \sqrt{2} \left[ c_{22} \ket{\xi_2} (c_{31} \bra{\xi_1} + c_{32} \bra{\xi_2}) + c_{23} c_{33}\ket{\xi_3}  \bra{\xi_3} \right] \ .
\end{split}
\ee
We have numerically checked that this operator faithfully reproduces the power spectral density obtained by the full operator $\hat b$ in the vicinity of the two-photon transition frequency.
We introduce similar operators for the two single-photon transitions by selecting those terms that oscillates at frequencies close to $\pm \alpha/2$. In particular, the operator which describes emission around the $g$-$e$ transition reads
\be
\begin{split}
\hat D_{ge}  = &\left(c_{11} \ket{\xi_1} +c_{12} \ket{\xi_2}\right) c_{23} \bra{\xi_3} \\
& +  \sqrt{2} \left( c_{21} \ket{\xi_1} + c_{22} \ket{\xi_2} \right) c_{33}\bra{\xi_3} \ .
\end{split}
\ee

\subsection{First and second-order correlation functions}

We compute first and second-order autocorrelations of a generic bosonic field $\hat F$ as
\begin{align}
g_1(\tau) &= \Trace [\hat F^\dagger e^{\mathcal L \tau}  \hat F \hat\rho_{\rm st} ] \ , \\
g_2(\tau) &= \Trace [\hat F^\dagger \hat F e^{\mathcal L \tau}  (\hat F \hat\rho_{\rm st} \hat F^\dagger)] \ ,
\end{align}
where $\rho_{\rm st}$ is the steady state solution of the master equation ($\mathcal L \hat\rho_{\rm st}=0$). We obtain the power spectral densities in Fig.~2(a-c) by Fourier transform of $g_1(\tau)$ for the full output field operator ($\hat F=\bout$). We obtain the correlation functions in Fig.~3(b,d) by calculating $g_2(\tau)$ for the $g$--$e$ doublet and triplet operators ($\hat F= \sqrt{\Gamma}\hat D_{ge}$ and $\sqrt{\Gamma}\hat T$, respectively).

\subsection{Analytic expressions}

We derive the following expressions for $n=3$, $\delta=0$, $\Gamma \ll \alpha,\Omega$ and $\epsilon \equiv \Omega/|\alpha| \ll 1$. In practice, they hold true for $\epsilon \lesssim 0.2$ [compare Fig.~4(a)].
The integrated photon flux at the two-photon transition is given by
\be
\begin{split}
P_{\rm TPRF} &= G_1(0)= \Gamma \mean{T^\dagger T} \\
&= \Gamma \frac{
\epsilon^2 \left[
55 \epsilon^4
-450 \epsilon^6
+4 \left(\Gamma/\alpha \right)^2
\right]
}{4 \left(\Gamma / \alpha \right)^2+9 \epsilon^4} \ ,    
\end{split}
\ee
and the zero-time photon autocorrelation is given by
\be
g_2(0)=\frac{\mean{T^\dagger T^\dagger T T}}{\mean{T^\dagger T}^2}=\frac{\left(4 \alpha ^2 \Gamma ^2+9 \Omega ^4\right) \left(4 \alpha ^2 \Gamma ^2+479 \Omega ^4\right)}{\left(4 \alpha ^2 \Gamma ^2+55 \Omega ^4\right)^2} \ .
\ee
The maximum value taken by $g_2(0)$ (for a resonant drive) is $g_2^{\rm max}(0)
\approx 2.831$ and is attained for the drive strength $\Omega_{g_2^{\rm max}}
\approx 1.730 \sqrt{|\alpha| \Gamma } $. In general, $g_2(0)$ takes larger values if the drive is blue-detuned, that is, $\delta>0$.

\section{Comparing measured correlation functions to theory}

\subsection{Effect of digital filtering}

When comparing our model to the experimental data in Fig.~3(a,b), we take the finite detection bandwidth into account by convolving the calculated $g_2(\tau)$ with the squared kernel
of the used digital filter twice. This is an approximation which holds in the limit of low signal-to-noise ratio relevant here; for more details, see Ref.~\cite{Lang2014}.

\subsection{Parameter values}

We obtain the theory traces in Fig.~3 by using independently measured sample parameters and the values $\Gamma/2\pi=2.2~\rm{MHz}$ [Fig.~3(a)] and $\Gamma/2\pi=3.5~\rm{MHz}$ [Fig.~3(b)] for the transmon decay rate. We use these as effective values resulting from the coupling of different transitions to a waveguide with an inhomogeneous density of states, a feature of our experimental setup that is not taken into account in our theory. In addition, we find that we obtain the best agreement to the data of Fig.~3(b) by assuming an additional detuning $\delta/2\pi=0.6~\rm{MHz}$ with respect to resonance, and a $0.2~\rm{dB}$ additional attenuation with respect to the calibrated value.

\section{Numerical analysis of photon correlations at finite detuning}

\begin{figure}
\includegraphics[width=\linewidth]{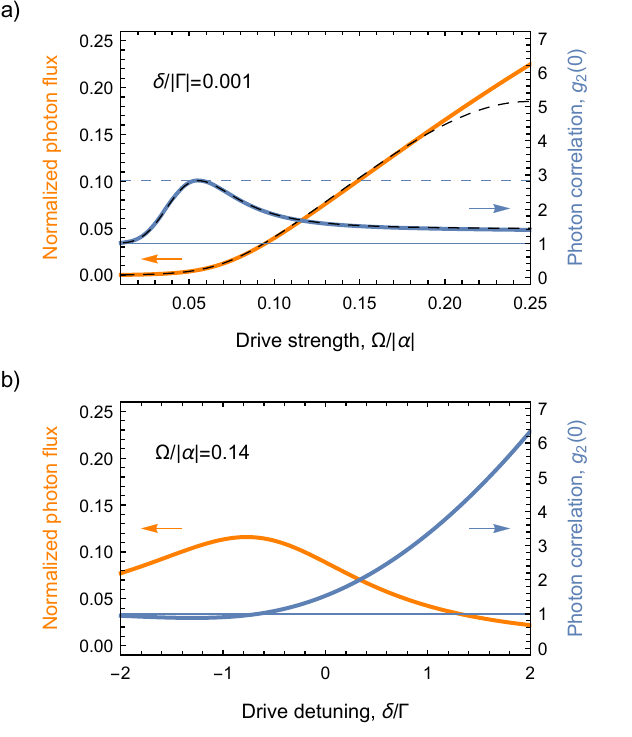}
\caption{\textbf{Calculated intensity and correlations for two-photon resonance fluorescence.} Photon flux normalized by the decay rate (orange, left axis) and normalized zero-time photon autocorrelation, $g_2(0)$ (blue, right axis) \textit{vs} (a) normalized drive strength, $\Omega/|\alpha|$, and (b) normalized drive detuning from the two-photon resonance, $\delta/\Gamma$. The anharmonicity is $\alpha/2\pi=-233~\rm{MHz}$ and the decay rate is $\Gamma/2\pi=2.5~\rm{MHz}$. The dashed curves in (a) are obtained using approximate analytic expressions (see text).
Horizontal lines indicate the asymptotic values $g_2(0)=1$ and $g_2(0)=2.831$.
}
\label{fig:TPRFtheory}
\end{figure}

We theoretically investigate how the intensity of the two-photon emission and its correlation properties depend on the drive strength, $\Omega$, and on the drive detuning from the two-photon resonance, $\delta=\omega_d-\omega_{gf}/2$ [Fig.~\ref{fig:TPRFtheory}(a) and \ref{fig:TPRFtheory}(b), respectively]. In the low-power limit, $\Omega \ll |\alpha|$, the two-photon signal is very weak compared to that from single-photon transitions.
This case has not been addressed in this work.
Emission becomes appreciable ($\simeq 0.1$ photons) when the ratio $\Omega/|\alpha|$ reaches the order of 0.1. In this regime, two-photon resonance fluorescence is assisted by the intermediate state $\ket{e}$, and displays superbunching. 

Detuning the drive frequency from the bare two-photon resonance
affects both the intensity and the correlation properties of the radiation emitted around the two-photon transition [Fig.~\ref{fig:TPRFtheory}(b)]. While the intensity is peaked at a slightly negative-detuned ($\delta \approx -\Gamma$) drive frequency, superbunching becomes more pronounced as the drive frequency is shifted towards higher frequencies. This increased superbunching is accompanied by the disappearance of the two farthest-detuned of the four single-photon peaks in the full emission spectrum [compare Fig.~2(a,c)]. This implies that single-photon relaxation primarily takes place via the cascade $\ket{-} \to \ket{\tilde{e}} \to \ket{-}$ [see the level scheme in Fig.~2(d)]. The steady-state solution of the master equation confirms that the dressed-state $\ket{-}$ is the most populated. This favors the two-photon process $\ket{-} \to \ket{\tilde{+}}$, giving rise to superbunching.

By contrast, for negative detunings the single-photon cascade $\ket{\tilde{+}} \to \ket{\tilde{e}} \to \ket{\tilde{+}}$ prevails and the system spends most of the time in $\ket{\tilde{+}}$ (and $\ket{\tilde{e}}$). This arrangement favors the resonant decay processes $\ket{\tilde{+}} \to \ket{\tilde{+}}$ and $\ket{\tilde{e}} \to \ket{\tilde{e}}$
, which do not entail population modulation of the state of the emitter. As a consequence, subsequent emission events are uncorrelated and the photon statistics is expected to be Poissonian ($g_2(0)=1$), in good agreement with the numerical result of Fig.~\ref{fig:TPRFtheory}(b).

\end{document}